\documentclass[runningheads]{llncs}
\usepackage{times}
\usepackage{epsfig}
\usepackage{graphicx}
\usepackage{amsmath}
\usepackage{amssymb}
\usepackage{nicefrac}       
\usepackage{url}            
\usepackage{booktabs}       
\usepackage{cite}
\usepackage{multirow}

\begin{document}
\title{Calibrating Histopathology Image Classifiers using Label Smoothing}
%
%
\vspace{-2mm}
\author{
Jerry Wei $^{1}$ \and
Lorenzo Torresani $^{1}$ \and
Jason Wei $^{1}$ \and
Saeed Hassanpour $^{1*}$}
\authorrunning{Wei et al.}
%
\institute{
Dartmouth College, Hanover, NH 03755, USA \\
${*}$ \url{saeed.hassanpour@darmouth.edu}}
\maketitle              

\vspace{-6mm}

\begin{abstract}
The classification of histopathology images fundamentally differs from traditional image classification tasks because histopathology images naturally exhibit a range of diagnostic features, resulting in a diverse range of annotator agreement levels.
However, examples with high annotator disagreement are often either assigned the majority label or discarded entirely when training histopathology image classifiers.
This widespread practice often yields classifiers that do not account for example difficulty and exhibit poor model calibration. 
In this paper, we ask: can we improve model calibration by endowing histopathology image classifiers with inductive biases about example difficulty?

We propose several label smoothing methods that utilize per-image annotator agreement. 
Though our methods are simple, we find that they substantially improve model calibration, while maintaining (or even improving) accuracy. 
For colorectal polyp classification, a common yet challenging task in gastrointestinal pathology, we find that our proposed agreement-aware label smoothing methods reduce calibration error by almost 70\%.
Moreover, we find that using model confidence as a proxy for annotator agreement also improves calibration and accuracy, suggesting that datasets without multiple annotators can still benefit from our proposed label smoothing methods via our proposed confidence-aware label smoothing methods.

Given the importance of calibration (especially in histopathology image analysis), the improvements from our proposed techniques merit further exploration and potential implementation in other histopathology image classification tasks.

\end{abstract}

\vspace{-10mm}
\section{Introduction}
\vspace{-3mm}

The success of modern deep learning paradigms on computer vision (CV) benchmarks such as ImageNet has led to widespread adoption of neural networks for medical image analysis \cite{Bulten2020}.
Convolutional neural networks, for instance, have been used for tasks such as lung cancer classification \cite{Coudray2017}, colorectal polyp classification \cite{Korbar2017}, and breast cancer detection \cite{Shen2019Deep}.
The nature of medical image classification tasks, however, differs substantially from standard benchmarks used in computer vision in two major ways.

First, on benchmark classification datasets such as ImageNet, CIFAR-10, and MNIST, most image labels are well-defined and have high annotator agreement (few examples exhibit high annotator disagreement). 
Labels for medical images, on the other hand, often have high annotator disagreement, especially for challenging tasks where diagnoses can differ substantially, even among expert clinicians \cite{Farris2008,Glatz2007,Warth1221}.
This is because the natural progression of diseases inherently creates a range of difficulty that has been observed in many medical image datasets \cite{Cheplygina2018Crowd,Wong2009}.
Despite this unique characteristic, existing literature typically uses the classic CV training procedure---using the majority vote of annotators as a one-hot encoded target \cite{Irvin2019,Chilamkurthy2018,Gulshan2016,Kanavati2020}.

The second way that medical image analysis differs from traditional computer vision is how models will be deployed in real-world settings. 
Whereas traditional CV applications such as facial recognition or self-driving cars are often used as standalone technologies, the initial adoption of medical image analysis models will likely be as artificial intelligence assistants for clinicians. 
A deep learning system could triage or provide second opinions in an image analysis pipeline, where images predicted to have unclear diagnoses are sent to separate queues for further manual inspection. 
In these scenarios, the confidence outputs of deep learning classifiers crucially influence clinician decisions, as high-confidence images will receive less attention and low-confidence images will receive more attention. 
In such settings, calibration is key.

\begin{figure*}[t]
    \centering
    \includegraphics[width=0.95\linewidth]{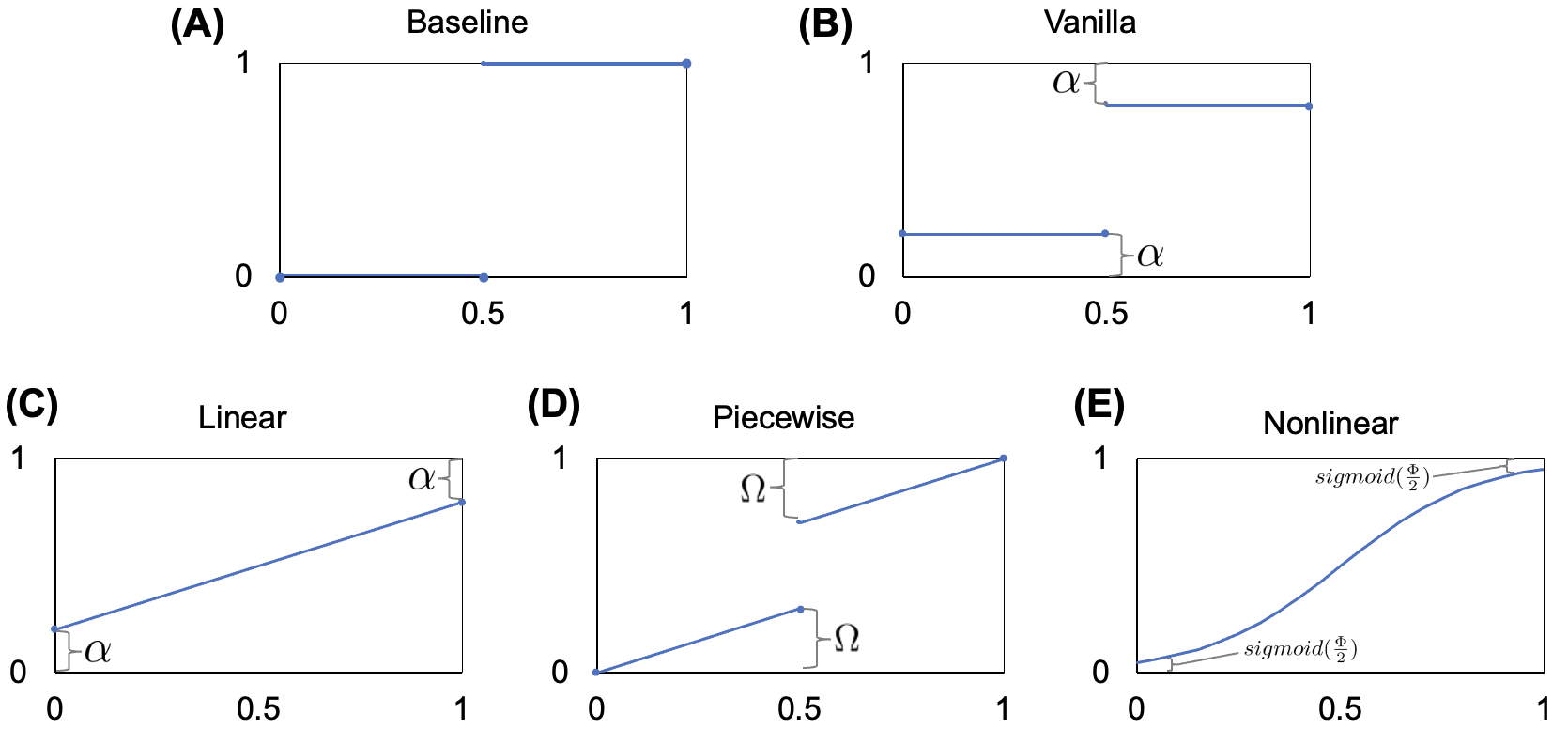}
    \caption{Proposed label smoothing methods. The $x$-axis represents annotator agreement level or model confidence, and the $y$-axis represents the label given to a particular example.}
    \vspace{-7mm}
    \label{fig:label_smoothing_visuals}
\end{figure*}

In this paper, we bring attention to applying label smoothing techniques for the classification of histopathology images, digitized images of surgical resection tissue that are typically examined under a microscope. 
Specifically, we leverage the intuition that many histopathology image datasets have annotations from multiple pathologists \cite{Chilamkurthy2018,Coudray2017,Bejnordi2017,Esteva2017,Irvin2019,Kanavati2020,Korbar2017} and operationalize these annotations into a technique that we call \textit{agreement-aware} label smoothing.
The soft targets in agreement-aware label smoothing depend on the annotator agreement for each individual image, providing the model with more information about the contents of the images. 
We also posit that model confidence can be used as a proxy for example difficulty in lieu of annotator agreement, allowing datasets that do not have annotations from multiple annotators to benefit from our proposed label smoothing methodology.
We thus propose an analogous \textit{confidence-aware} label smoothing technique that closely resembles agreement-aware label smoothing but substitutes annotator agreement data with model confidence outputs. 

We conduct empirical experiments on colorectal polyp classification---one of the highest volume tasks in clinical pathology.
We find that both agreement-aware label smoothing and confidence-aware label smoothing improve calibration more than vanilla label smoothing and hard-target models do, while maintaining or improving accuracy.

\vspace{-3mm}
\section{Label Smoothing}
\vspace{-3mm}

Label smoothing \cite{muller2020does} is a paradigm proposing to ``smooth" labels by encouraging small logit gaps, addressing the issue of traditional one-hot encoded labels resulting in models being less-calibrated and too overconfident in their predictions.
In practice, vanilla label smoothing assigns a smooth label $y$ using a hyperparameter $\alpha$ $\in$ $(0, 1]$, one-hot target $y_k$, and the total number of classes $K$, as depicted in Figure \ref{fig:label_smoothing_visuals}B and formalized below.

\begin{equation}
    y = (1 - \alpha) * y_k + \frac{\alpha}{K}\ .
    \label{eq:vanilla_label_smoothing}
\end{equation}

\vspace{-3mm}
\subsection{Agreement-Aware Label Smoothing}
\vspace{-2mm}

Because histopathology image classification datasets often contain annotator agreement data that could potentially be used for more-precise label smoothing \cite{Chilamkurthy2018,Coudray2017,Bejnordi2017,Esteva2017,Irvin2019,Kanavati2020,Korbar2017} , we propose three types of agreement-aware label smoothing paradigms.
We first implement a linear agreement-aware label smoothing method (which closely matches vanilla label smoothing) in order to examine the sole effect of including annotator agreement data without changing the format of vanilla label smoothing.
This method simply replaces the one-hot label in vanilla label smoothing with $n_k$---the number of annotators that labeled a given example as class $k$---divided by the total number of annotators $N$ (Equation \ref{eq:linear_annotator_label_smoothing}).
Our linear agreement-aware method continues to use hyperparameter $\alpha$ $\in$ $(0, 1]$.
A visual representation of this method is shown in Panel C in Figure \ref{fig:label_smoothing_visuals}.

\begin{equation}
    y = (1 - \alpha) * \frac{n_k}{N} + \frac{\alpha}{K}\ .
    \label{eq:linear_annotator_label_smoothing}
\end{equation}

We also explore whether implementing a variable-sized discontinuity in the vanilla label smoothing equation can improve the precision of smoothed targets.
Thus, we implement a piecewise agreement-aware label smoothing method (Equation \ref{eq:piecewise_annotator_label_smoothing}) which uses a different hyperparameter $\Omega$ $\in$ $(0, 0.5]$.
In this system of equations, we define the number of annotators needed for a majority $n_m = \left\lceil\frac{N}{K}\right\rceil$.
This method can be seen in Panel D in Figure \ref{fig:label_smoothing_visuals}.

\begin{equation}
    \begin{cases} 
    y = (1 - \Omega) + \Omega(\frac{n_k - n_m}{n_m - 1}), & \mbox{if } n_k > n_m \\
    y = 0.5, & \mbox{if } n_k = n_m\\
    y = \Omega(\frac{n_k}{n_m - 1}), & \mbox{if } n_k < n_m
    \end{cases}\ .
\label{eq:piecewise_annotator_label_smoothing}
\end{equation}

Finally, we address whether more-heavily penalizing images with higher disagreement can produce better-calibrated models.
In Equation \ref{eq:nonlinear_annotator_label_smoothing}, we define a nonlinear agreement-aware label smoothing method using hyperparameter $\Phi > 0$. 
For this nonlinear function, we use $f(x) = sigmoid(x)$; the nonlinearity of this function results in the targets of images being more-heavily penalized as annotator disagreement increases. 
The visual representation of this equation is shown in Panel E in Figure \ref{fig:label_smoothing_visuals}.

\begin{equation}
    y = f(\Phi(\frac{n_k}{N} - \frac{1}{2}))\ .
\label{eq:nonlinear_annotator_label_smoothing}
\end{equation}

We model our agreement-aware label smoothing equations for binary classification tasks, though our equations could be easily adjusted for other scenarios.
For example, the nonlinear function can be easily changed to $f(x) = softmax(x)$ if $K > 2$. 

\vspace{-4mm}
\subsection{Confidence-Aware Label Smoothing}
\label{section:confidence-aware-label-smoothing}
\vspace{-2mm}

Many datasets do not have annotation data from multiple annotators, and so a natural question is whether these datasets can still benefit from our proposed label smoothing methods.
To answer this question, we use the confidence outputs of a baseline model (which can be obtained regardless of the number of annotators a dataset has) as a proxy for example difficulty.
This method thus requires training the model twice---first without label smoothing in order to obtain model confidence scores for each training example, then again with our confidence-aware label smoothing methods.

While prior work has focused mostly on using model confidence as target labels \cite{hinton2015distilling,papernot2016distillation,wei2020learn}, the novelty of confidence-aware label smoothing lies in its fusion of the idea of using model confidence as target labels with our proposed agreement-aware label smoothing formulas.
Thus, for these methods, we replace annotator agreement level $\frac{n_k}{N}$ in our proposed agreement-aware label smoothing equations with the confidence value $c_k$ for a given example to obtain analagous confidence-aware label smoothing equations.
This does not change the overall shape of the function, and confidence-aware counterparts for agreement-aware label smoothing methods are still represented by their respective panels in Figure \ref{fig:label_smoothing_visuals}.
We first define vanilla confidence-aware label smoothing in this manner, obtaining Equation \ref{eq:vanilla_confidence_label_smoothing} from Equation \ref{eq:vanilla_label_smoothing} by substituting model confidence for annotator agreement level.

\begin{equation}
    y = (1 - \alpha) * \left\lfloor{c_k}\right\rceil + \frac{\alpha}{K}\ .
    \label{eq:vanilla_confidence_label_smoothing}
\end{equation}

Similarly, for linear and nonlinear confidence-aware label smoothing, we simply replace $\frac{n_k}{N}$ with $c_k$ in Equations \ref{eq:linear_annotator_label_smoothing} and \ref{eq:nonlinear_annotator_label_smoothing}, respectively. 
Piecewise confidence-aware label smoothing, however, is more complicated and requires additional modifications as a result of the use of $n_m$, so we show the equation for this separately in Equation \ref{eq:piecewise_confidence_label_smoothing}.

\begin{equation}
    \begin{cases} 
    y = (1 - \Omega) + (\frac{c_k - 0.5}{0.5}) * \Omega, & \mbox{if } c_k > 0.5 \\
    y = 0.5, & \mbox{if } c_k = 0.5\\
    y = (\frac{c_k}{0.5}) * \Omega, & \mbox{if } c_k < 0.5
    \end{cases}\ .
\label{eq:piecewise_confidence_label_smoothing}
\end{equation}

\vspace{-4mm}
\section{Experimental Setup}
\vspace{-3mm}

For our experiments, we use the Minimalist Histopathology Image Analysis Dataset (MHIST) \cite{wei2021petri}, a publicly-available dataset for the classification of colorectal polyps.
MHIST focuses on the clinically-important binary classification between hyperplastic polyps (HPs) and sessile serrated adenomas (SSAs), a challenging yet common diagnostic distinction \cite{Farris2008,Glatz2007,Wong2009}. 
MHIST contains a total of 3,152 images (2,175 for training and 977 for evaluation), each annotated with a binary label of either HP or SSA. 
MHIST includes independently-annotated labels from seven gastrointestinal pathologists---we leverage this annotator agreement data to endow image classifiers with inductive biases about example difficulty.

We also use the ResNet architecture \cite{He2015}, a common choice for classifying histopathology images. 
We use the same hyperparameters and training/evaluation sets as implemented in the original MHIST paper \cite{wei2021petri}.
We measure model accuracy using the area under the receiver operating characteristic curve (AUC; higher is better) and also measure model calibration using expected calibration error \cite{Naeini2015Obtaining} (ECE; lower is better), the weighted average over the difference between accuracy and confidence computed over a given number of bins (for our experiments, we use 15 bins). 
ECE is best when a model's accuracy and confidence are identical (e.g., the model obtains 90\% accuracy on images that it predicts with 90\% confidence).
For each model, we report the mean and standard deviation of these AUC and ECE values calculated over 10 different seeds.

\vspace{-2mm}
\section{Label Smoothing: Annotator Agreement}
\vspace{-2mm}
\subsection{Improving Accuracy and Calibration}
\vspace{-2mm}

\begin{table}[t!]
    \setlength{\tabcolsep}{5pt}
    \centering
    \small
    \begin{tabular}{l c c}
        \toprule
        Model & AUC & ECE \\
        \midrule
        Baseline & 84.7 $\pm$ 0.8 & 8.9 $\pm$ 1.4\\
        Vanilla Label Smoothing & 85.6 $\pm$ 0.6 & 3.2 $\pm$ 0.6\\
        Agreement-Aware (Linear) & 86.1 $\pm$ 0.9 & 5.3 $\pm$ 0.7\\
        Agreement-Aware (Piecewise) & \textbf{86.4} $\pm$ 0.4 & 2.9 $\pm$ 0.4\\
        Agreement-Aware (Nonlinear) & 86.3 $\pm$ 0.7 & \textbf{2.8} $\pm$ 0.6\\
        \bottomrule
    \end{tabular}
    \vspace{2mm}
    \caption{Agreement-aware label smoothing improves performance (higher AUC) and model calibration (lower ECE). Means and standard deviations are reported across 10 seeds.}
    \label{tab:summary_agreement_aware}
    \vspace{-5mm}
\end{table}

\begin{figure*}[t]
    \centering
    \includegraphics[width=0.9\linewidth]{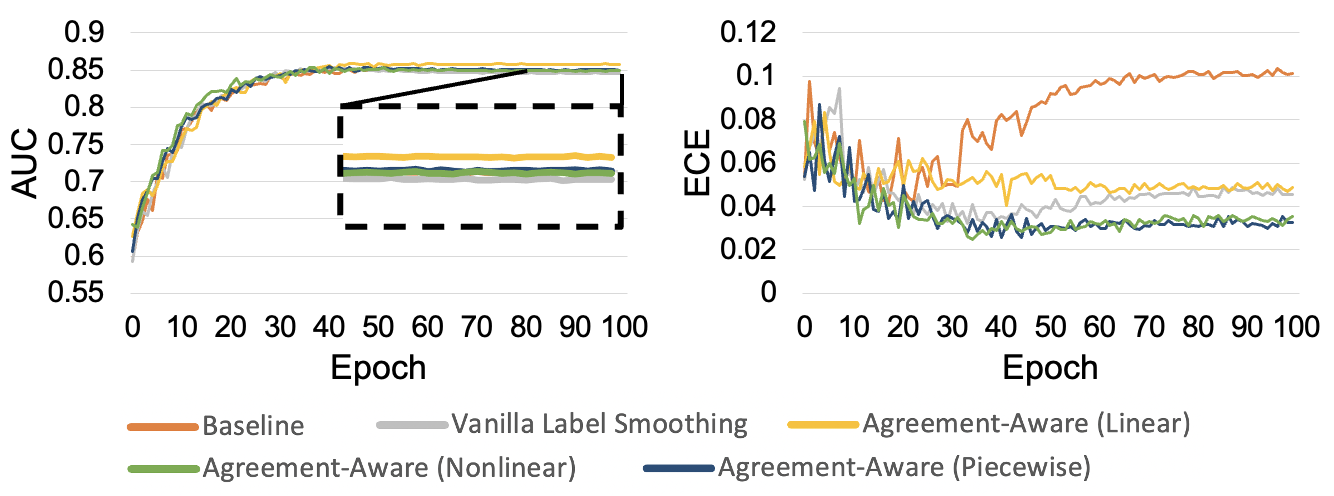}
    \caption{Whereas the baseline model becomes uncalibrated after training for $\sim$30 epochs (ECE goes up), models trained using agreement-aware label smoothing remained well-calibrated throughout training (ECE stays constant). The models trained using label smoothing achieved similar or better performance (AUC) than the baseline.}
    \vspace{-5mm}
    \label{fig:auc_ece_epochs_annotator}
\end{figure*}

In Table \ref{tab:summary_agreement_aware}, we summarize AUC and ECE values for the best-performing models (which we define as the model with the lowest ECE among models that have a higher AUC than the baseline model) for each label smoothing method.
We find that our agreement-aware label smoothing methods decrease mean ECE by up to 6.1 percentage points (a 68.5\% decrease) and also improve mean AUC by up to 1.7\%, suggesting that our methods can improve calibration without sacrificing accuracy.

We further analyze these models in Figure  \ref{fig:auc_ece_epochs_annotator} by plotting the AUC and ECE of these models on our testing set throughout training. 
We find that, as expected, the baseline model first experiences a decrease in ECE, then an increase as it begins to overfit to the training set.
For models trained with label smoothing, however, the ECE decreases and does not experience a dramatic increase during later epochs.
Instead, the ECE holds relatively constantly once it reaches its lowest values.
We also found that vanilla label smoothing and linear agreement-aware label smoothing both significantly improved upon the baseline model in terms of ECE.
Additionally, piecewise agreement-aware label smoothing and nonlinear agreement-aware label smoothing seem to further improve upon these two methods and achieved the best calibration among models trained using our label smoothing methods.

\vspace{-5mm}
\subsection{Hyperparameter Selection}
\vspace{-2mm}

\begin{figure*}[t]
    \centering
    \includegraphics[width=\linewidth]{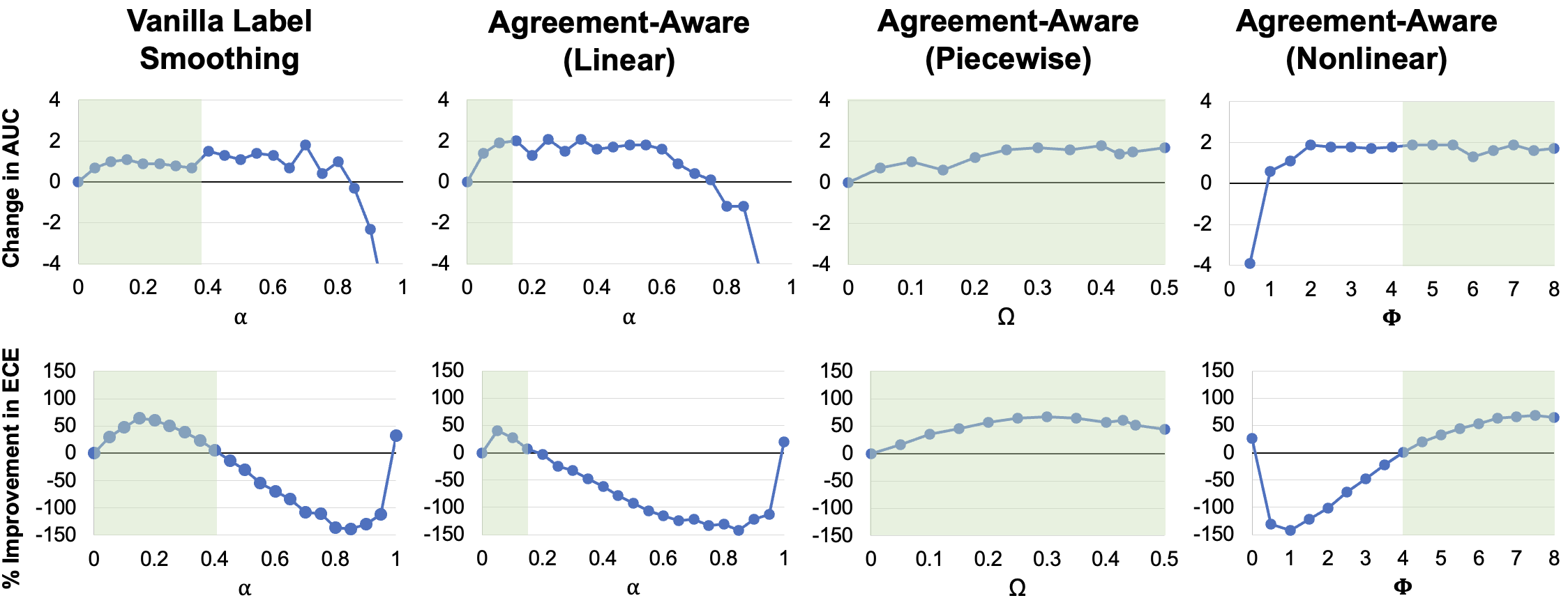}
    \caption{Top: change in AUC from the baseline model. Bottom: percent improvement in ECE from the baseline model. The highlighted green area indicates the hyperparameter range where both AUC is higher and ECE is lower, compared with the baseline model.}
    \vspace{-5mm}
    \label{fig:auc_ece_improvement_annotator}
\end{figure*}

We conduct extensive ablation studies for each of our label smoothing methods to analyze how robust our models are to different hyperparameter selections and to gain intution on how different strengths of our agreement-aware label smoothing methods will affect overall performance.
In Figure \ref{fig:auc_ece_improvement_annotator}, we show increases/decreases in AUC and ECE on the testing set between the baseline model and models trained using each of the label smoothing methods at many hyperparameter values.
We find that all label smoothing methods are able to improve both AUC and ECE for some hyperparameter ranges, and the greatest improvement in calibration (a 68.5\% decrease in ECE) occurs when using nonlinear agreement-aware label smoothing with $\Phi = 7.5$.
Generally, we find that piecewise agreement-aware label smoothing and nonlinear agreement-aware label smoothing are able to achieve the greatest improvements in model calibration, though vanilla label smoothing and linear agreement-aware label smoothing still yield significant improvements.

\vspace{-4mm}
\section{Label Smoothing: Model Confidence}
\vspace{-3mm}

Because many datasets do not include data from multiple annotators, in this section we conduct experiments on whether our proposed label smoothing methods can still benefit datasets with only one label per example.
To do so, we use model confidence in lieu of annotator agreement as a proxy for example difficulty, as proposed in Section \ref{section:confidence-aware-label-smoothing}.
We obtain model confidence data by using each image's confidence output from our baseline model (which was only trained using binary gold standard annotations such as what would be available for a dataset with only one annotator).
We then use these confidence outputs to conduct our experiments by using these them to replace the annotator agreement data used in Equations \ref{eq:vanilla_label_smoothing} -- \ref{eq:nonlinear_annotator_label_smoothing}.

\begin{table}[t]
    \setlength{\tabcolsep}{5pt}
    \centering
    \small
    \begin{tabular}{l c c}
        \toprule
        Model & AUC & ECE \\
        \midrule
        Baseline & 84.7 $\pm$ 0.8 & 8.9 $\pm$ 1.4\\
        Confidence-Aware (Vanilla) & 85.2 $\pm$ 0.8 & 3.6 $\pm$ 0.5\\
        Confidence-Aware (Linear) & 85.9 $\pm$ 0.8 & 3.5 $\pm$ 0.6\\
        Confidence-Aware (Piecewise) & 85.4 $\pm$ 0.7 & 8.4 $\pm$ 1.4\\
        Confidence-Aware (Nonlinear) & \textbf{86.2} $\pm$ 1.1 & \textbf{3.2} $\pm$ 0.4\\
        \bottomrule
    \end{tabular}
    \vspace{2mm}
    \caption{Confidence-aware label smoothing improves performance (higher AUC) and model calibration (lower ECE). Means and standard deviations are reported across 10 seeds.}
    \label{tab:summary_confidence_aware}
    \vspace{-7mm}
\end{table}

\vspace{-3mm}
\subsection{Improving Accuracy and Calibration}
\vspace{-2mm}

We report AUC and ECE values on our testing set for the best-performing models (defined as the model with the lowest ECE among models with a higher AUC than the baseline) for each of our confidence-aware label smoothing methods in Table \ref{tab:summary_confidence_aware}.
We find that, although not as effective as agreement-aware label smoothing, confidence-aware label smoothing is still very effective in improving both accuracy and calibration.
Confidence-aware label smoothing methods were able to increase AUC by up to 1.5\% and decrease ECE by 5.7 percentage points (an improvement of 64.0\%).
We found, however, that piecewise confidence-aware label smoothing was ineffective, likely because it was overly-complex for already-sophisticated data such as confidence values, resulting in targets that were too convoluted to extract any new information from.


\vspace{-3mm}
\subsection{Hyperparameter Selection}
\vspace{-2mm}

\begin{figure*}[t]
    \centering
    \includegraphics[width=\linewidth]{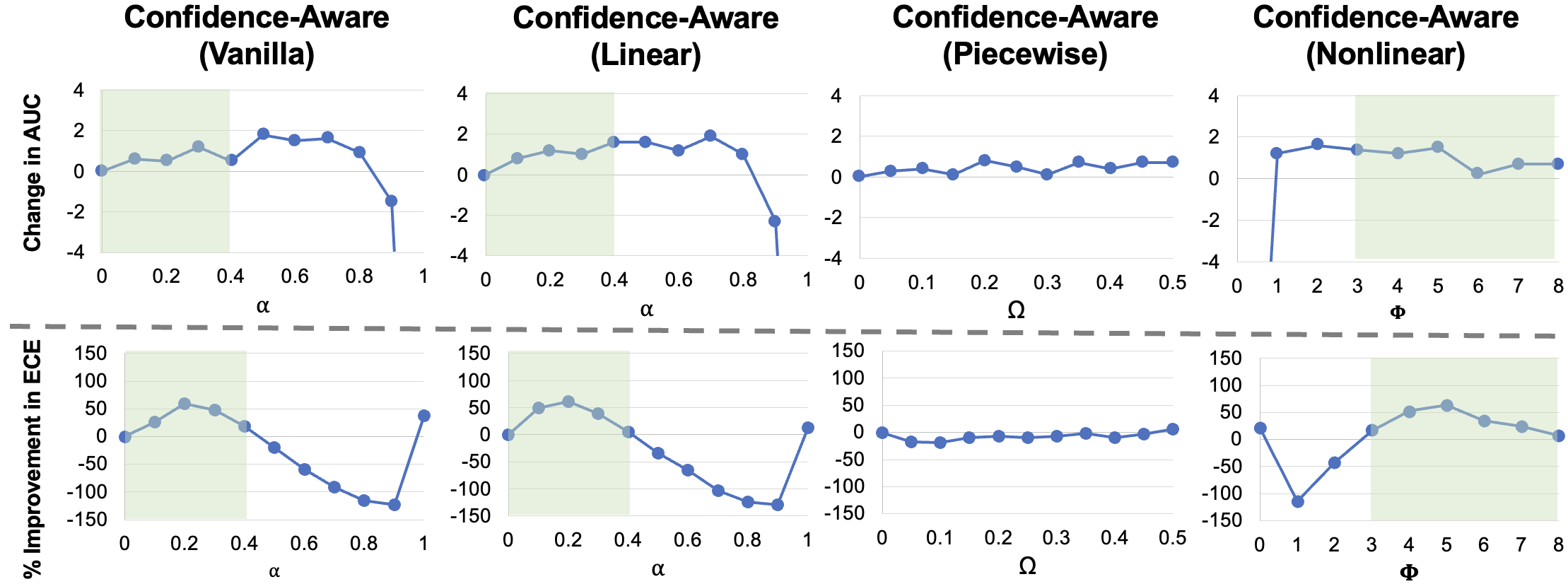}
    \caption{Top: change in AUC from the baseline model. Bottom: percent improvement in ECE from the baseline model. The highlighted green area indicates the hyperparameter range where both AUC is higher and ECE is lower, compared with the baseline model.}
    \vspace{-5mm}
    \label{fig:auc_ece_improvement_confidence}
\end{figure*}

We once again study how robust our confidence-aware label smoothing methods are in terms of choosing a reasonable hyperparameter, and we analyze how different strengths of our confidence-aware label smoothing methods affect model performance.
In Figure \ref{fig:auc_ece_improvement_confidence}, we show changes in AUC and improvements in ECE for our confidence-aware label smoothing methods.
We find that $0 < \alpha < 0.4$ improves model calibration for vanilla and linear confidence-aware label smoothing, and $3 < \Phi < 8$ improves calibration for nonlinear confidence-aware label smoothing. 
For piecewise confidence-aware label smoothing, however, model calibration only improves at $\Omega = 0.5$, suggesting that piecewise confidence-aware label smoothing is not robust to hyperparameter selection and offers the least improvement out of all confidence-aware label smoothing methods.

\vspace{-3mm}
\section{Related Work and Discussion}
\vspace{-2mm}
Prior work has found that modern neural networks, though accurate, can be poorly-calibrated \cite{muller2020does,guo2017calibration}. 
However, calibration is a critical factor to account for when training models (especially for applications such as self-driving cars and health care) and allows for better model interpretability \cite{guo2017calibration}.
Label smoothing, originally proposed by Muller et al. \cite{muller2020does}, seeks to improve calibration and has been found to reduce expected calibration error on computer vision benchmarks such as ImageNet and CIFAR10.

In the domain of medical image analysis, Pham et al. \cite{pham2020interpreting} proposed to remap targets to random numbers close to one, finding that this method improved model performance on the CheXpert Dataset \cite{Irvin2019} by approximately 1.4\%.
Moreover, Xi et al. \cite{Xi2019weakly} addressed uncertainties of correctness using a spatial label smoothing technique to reduce the need for well-annotated data while still achieving satisfactory performance.

Previous research has been largely focused purely on improving calibration and accuracy using label smoothing techniques and primarily uses one-hot encoded labels to do so.
However, the existing literature has disregarded the vast amount of information that lies in example difficulty, which could potentially be useful for more-precise label smoothing.
Our work thus seeks to expand on existing label smoothing methods by incorporating example difficulty.
Perhaps most related to our work, Peterson et al. \cite{Peterson2019} found that performance on CIFAR10 can be improved by sampling ground truth labels from a distribution of human annotations, which is equivalent to our linear agreement-aware method.
Our work proposes a superset of methods operationalized via label smoothing.

Although we intentionally chose the colorectal polyp classification task because it is a common and diagnostically-challenging problem in gastrointestinal pathology---the most-relevant dataset with multiple annotations that we have access to at this time---our study nevertheless is only evaluated on one dataset. 
As such, we do not consider our encouraging empirical results as validation of our label smoothing methods for all histopathology tasks.
Instead, our paper has questioned the traditional practice of assigning hard labels for histopathology image classification and proposed a simple yet effective alternative that invites further research in this direction.

In this paper, we have proposed two well-motivated sets of label smoothing methods for improving the calibration and accuracy of histopathology image classifiers: \textit{agreement-aware} label smoothing, which uses annotator agreement data, and \textit{confidence-aware} label smoothing, which uses model confidence.
We apply our method to a colorectal polyp classification task, finding that agreement-aware label smoothing can reduce calibration error by 68.5\% while increasing AUC by up to 1.7\%, and confidence-aware label smoothing can reduce calibration error by 64.0\% while increasing AUC by up to 1.5\%.
Our paper aims to demonstrate the potential usefulness of including example difficulty when implementing label smoothing, and we hope that our methods can be further implemented across the field of histopathology image classification due to its low computational cost and large potential in improving model calibration and performance.

\clearpage
{\small
\bibliographystyle{splncs04}
\bibliography{egbib}
}

\end{document}